\documentclass[superscriptaddress,amsmath,amssymb,aps,reprint,prb,twocolumn]{revtex4-2}

\usepackage{graphicx}
\usepackage{dcolumn}
\usepackage{bm}
\usepackage{hyperref}
\usepackage[normalem]{ulem}
\hypersetup{
	citecolor = blue,
	colorlinks = true,
	urlcolor = blue
}

\usepackage{xcolor}
\usepackage{braket}

\begin{document}


\title{\textbf{Torsional oscillation of carbon nanotubes driven by electron spins}}

\author{Koji Yamada}
\affiliation{Institute for Solid State Physics, University of Tokyo, Kashiwa 277-8581, Japan}

\author{Wataru Izumida}
\affiliation{Department of Physics, Tohoku University, Sendai 980-8578, Japan}

\author{Mamoru Matsuo}
\affiliation{Kavli Institute for Theoretical Sciences, University of Chinese Academy of Sciences, Beijing, 100190, China.}
\affiliation{CAS Center for Excellence in Topological Quantum Computation, University of Chinese Academy of Sciences, Beijing 100190, China}
\affiliation{Advanced Science Research Center, Japan Atomic Energy Agency, Tokai, 319-1195, Japan}
\affiliation{RIKEN Center for Emergent Matter Science (CEMS), Wako, Saitama 351-0198, Japan}

\author{Takeo Kato}
\affiliation{Institute for Solid State Physics, University of Tokyo, Kashiwa 277-8581, Japan}

\date{\today}

\begin{abstract}
We theoretically investigate the current-induced excitation of torsional vibrations in a suspended carbon nanotube (CNT) quantum dot. By considering a CNT clamped between half-metallic ferromagnetic electrodes with an antiparallel magnetization configuration, we demonstrate that the spin-rotation coupling (SRC) enables the transfer of angular momentum from electron spins to the mechanical torsional mode under a constant source-drain voltage. Using a master-equation approach to analyze the coupled dynamics of the dot levels and a quantized torsional oscillator, we evaluate the steady-state current and phonon distribution. We find that when the Zeeman splitting matches the torsional phonon energy, the system exhibits sharp resonant behavior in the current, accompanied by a significant increase in the phonon population. Our estimates for realistic device parameters indicate that this spin-driven mechanism can drive CNT torsional vibrations with detectable amplitudes. 
This work provides a theoretical basis for resonant current-controlled actuation of CNT torsional modes via SRC-mediated transfer of spin angular momentum.
\end{abstract}

\maketitle

\section{Introduction}
\label{sec:introduction}

Carbon nanotubes (CNTs) form a standard platform for nanoelectromechanical systems (NEMS) because a single suspended tube combines a small mass with a large stiffness and can reach high resonance frequencies~\cite{Craighead2000,Sapmaz2003,Sazonova2004,PootZant2012}.
While carbon nanotubes played a central role in early studies of NEMS, recent advances have extended NEMS research to a variety of nanomaterials~\cite{Xu:2022aa}.
Beyond flexural motion, torsional degrees of freedom of CNT devices have also been resolved, including torsion springs and torsional resonators~\cite{Papadakis2004,Hall2008,Mori2020}. 
A suspended CNT simultaneously acts as a quantum dot, so that electronic transport and a selected mechanical mode can be treated within a single nanoscale device~\cite{Huttel2009,Steele2009,Lassagne2009,Laird2015,Izumida2015,Izumida2016}. 
This setting invites actuation schemes in which the mechanical variable is driven by a microscopic channel of angular-momentum transfer rather than by an electrostatic force.

Electrical actuation in NEMS is often based on capacitive forces and therefore couples most directly to displacement~\cite{Ekinci2005}. This coupling fits naturally with flexural modes, where the generalized force is a transverse force. The torsional mode, by contrast, couples to a torque about the tube axis. Although a controlled torque can be generated using torsional nanomechanical devices~\cite{Papadakis2004,Hall2008,Mori2020}, this approach constrains device geometry and electrode design.
A torque-based drive is nonetheless of interest because torsional motion probes mechanical angular momentum directly and provides a route to couple nanomechanics to the internal angular momentum carried by electrons.

A microscopic source of torque is provided by the conversion between spin angular momentum and mechanical rotation. 
The classical gyromagnetic effects, associated with the Barnett and Einstein--de Haas effects~\cite{Barnett1915,Einstein1915,Barnett1935,Scott1962}, express the exchange of angular momentum between magnetization and rotation.
The gyromagnetic effects have attracted renewed attention in the 21st century in the context of spin current induced by inhomogeneities of the angular velocity~\cite{Matsuo2013,Matsuo2017,Matsuo2017b,Takahashi2016,Takahashi2020,Tabaei2020,Tabaei2021,Kobayashi2017,Kurimune2020,Kurimune2020b,Tateno2021,Funato2025}.
They have also been utilized for mechanical torsion generation using electron spin relaxation on nanoscale objects~\cite{Cleland2003book,Mohanty2004,Wallis2006,Zolfagharkhani2008,Harii2019}.
In a quantum description, the coupling between a rotating frame and an electron spin is given by the spin-rotation coupling (SRC)~\cite{Hehl1990,Matsuo2011}
\begin{align}
\hat{H}_{\mathrm{SR}} = - \boldsymbol{\omega}\cdot \hat{\boldsymbol{S}},
\label{eq:HSR}
\end{align}
where $\boldsymbol{\omega}$ is the angular velocity and $\hat{\boldsymbol{S}}$ is the spin operator. Equation~(\ref{eq:HSR}) implies that a spin flip can act as a torque impulse on a mechanical coordinate that carries angular momentum. 
In CNTs, where twisting motion is well defined and can be quantized~\cite{Suzuura2002}, SRC-mediated spin-phonon conversion has been discussed in several contexts~\cite{Hamada2018}.

A recent example of SRC-based actuation is the Einstein--de Haas nanorotor proposed by Izumida et al.~\cite{Izumida2022}. In that setup, a CNT quantum dot is embedded between ferromagnetic electrodes with antiparallel magnetizations. Under steady bias voltage, a current requires spin-flip processes on the dot, and each spin flip transfers angular momentum to the mechanical rotation through Eq.~(\ref{eq:HSR}). 
Reference~\cite{Izumida2022} demonstrated continuous rigid-body rotation induced by this angular momentum transfer and discussed the efficiency of conversion from electron spins to rotation of the CNT.
In that nanorotor setup, the relevant mechanical coordinate is the continuous rigid-body rotation of the entire CNT, so the transferred spin angular momentum is accumulated as mechanical angular momentum. By contrast, once the CNT is clamped at both ends, the rotational degree of freedom is replaced by discrete elastic torsional eigenmodes, and the spin transfer is governed by an energy-selective resonance rather than by unrestricted rotational acceleration.

The present work considers a complementary regime in which the mechanical degree of freedom is not rigid-body rotation but the fundamental torsional mode of a suspended CNT clamped between two electrodes. The clamped geometry removes the need to engineer a bearing for a freely rotating shaft while still allowing a well-defined mechanical angular velocity associated with torsional vibration. 
The device is again operated in a spin-selective transport configuration using ferromagnetic electrodes with antiparallel magnetizations. 
For simplicity, we consider fully spin-polarized ferromagnetic (half-metallic) electrodes in the main calculation, while partially spin-polarized electrodes are discussed in Sec.~\ref{sec:partial}.
A steady current is then limited by the spin valve effect, and transport proceeds through spin flips on the dot. Because SRC couples each spin-flip event to the exchange of mechanical angular momentum, electron transport can pump the torsional oscillator by creating (and absorbing) torsional phonons. The pumping is most efficient when the Zeeman splitting $h$ is tuned close to the torsional phonon energy, $h \simeq \hbar \omega_{0}$, within the effective linewidth of the spin-phonon resonance, where $\omega_0$ is the eigenfrequency of the fundamental torsional mode.
This produces a sharp transport signature: the steady-state current exhibits a ridge along the resonance line, accompanied by a nonequilibrium increase in the phonon population.

We analyze this mechanism using a master equation approach for the coupled dynamics of the dot and a quantized torsional mode. The calculation yields the steady-state phonon distribution and the current as functions of bias voltage, magnetic field, temperature, and the relaxation rate of the torsional mode. For parameter values consistent with suspended CNT devices, the resulting torsional angle at the center of the tube reaches an angle of order one degree in the driven steady state. 
Our study thus provides a resonant torsional mode extension of SRC-mediated Einstein--de Haas actuation in a CNT nanoelectromechanical system. 
Unlike the nanorotor scenario of Ref.~\cite{Izumida2022}, the present setup converts spin transfer into resonant torsional-phonon pumping rather than continuous rotation. 
This mechanism offers a route to torque actuation of CNT torsional modes by spin angular momentum, distinct from force-based electrical drives that mainly address flexural motion.

The remainder of the paper is organized as follows. Section~\ref{sec:model} introduces the model Hamiltonian and the SRC-mediated spin-phonon coupling for a clamped CNT torsional mode. Section~\ref{sec:method} presents the master equation and the transition rates. Section~\ref{sec:results} discusses the steady-state current and the phonon distribution, including the resonance and the dependence on relaxation and temperature. Section~\ref{sec:discussion} comments on the effects of realistic ferromagnetic electrodes with finite spin polarization, as well as extensions to realistic CNT spectra with valley degrees of freedom and spin-orbit interaction.
Section~\ref{sec:summary} summarizes the results.

\section{Model}\label{sec:model}

\begin{figure}[t]
\centering
\includegraphics[width=8.5cm]{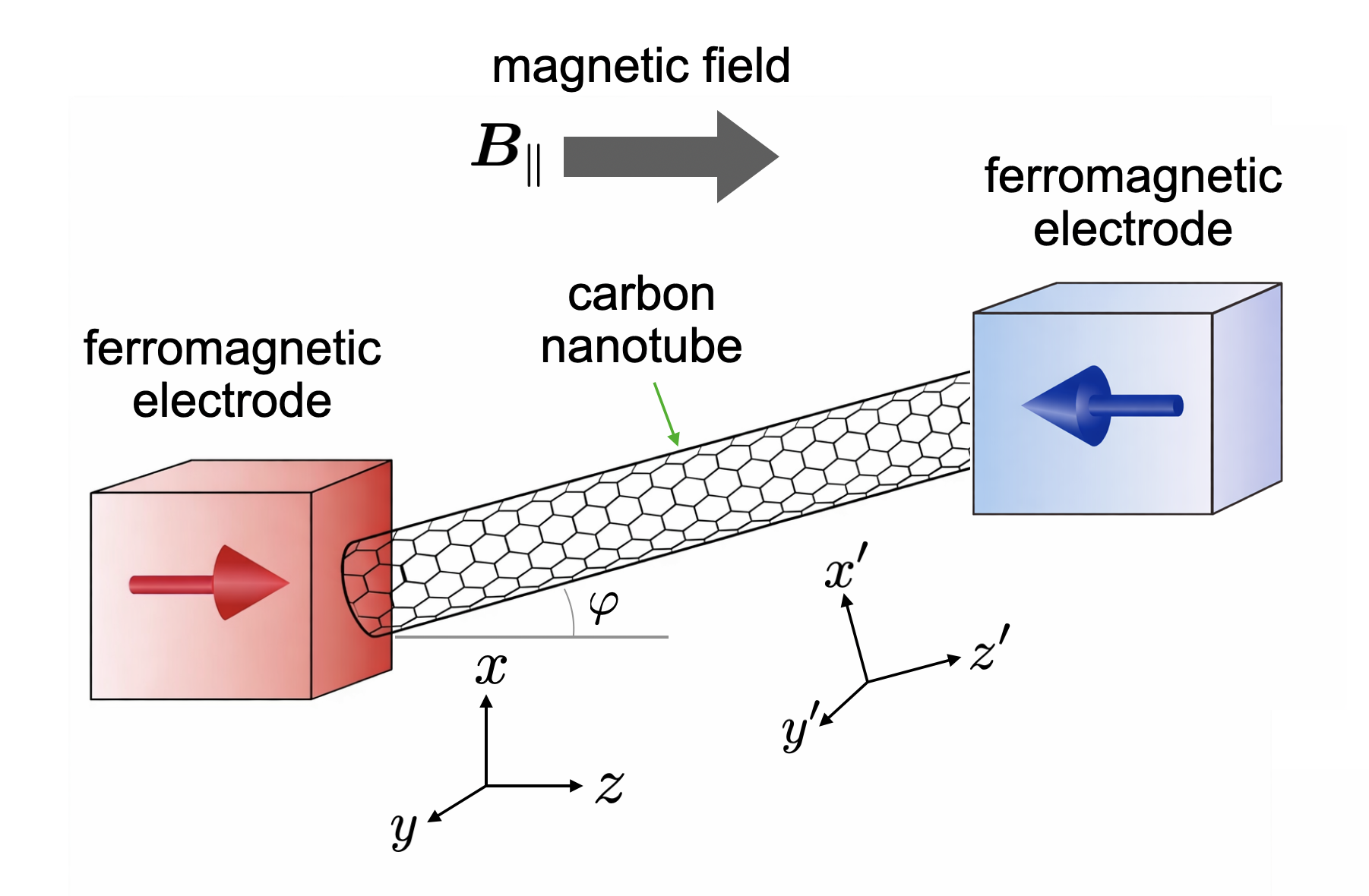}
\caption{Schematic illustration of the model: A suspended CNT of length \( L \) is connected to half-metallic ferromagnetic electrodes magnetized along opposite directions.}
\label{fig:model}
\end{figure}

We consider a suspended single-wall CNT of length \( L \), whose ends are rigidly clamped by ferromagnetic electrodes, as illustrated in Fig.~\ref{fig:model}.
The left and right electrodes are assumed to be half-metallic, magnetized along the \( +z \) and \( -z \) directions, respectively.
For simplicity, we assume ideal antiparallel alignment of the lead magnetizations in the main model. A slight misalignment may produce a finite background leakage current, analogous to the case of partially spin-polarized electrodes discussed in Sec.~\ref{sec:partial}.
A static magnetic field \( {\bm B} \parallel \hat{z} \) is applied.
The axis of the CNT is tilted by an angle $\varphi$ with respect to the magnetic-field direction.
The system Hamiltonian is given by 
\begin{align}
H &= H_{\mathrm{dot}} + H_{\mathrm{lead}} + H_{\mathrm{lead\text{-}dot}} \notag \\
& \hspace{5mm} + H_{\mathrm{ph}} + H_{\mathrm{env}} + H_{\mathrm{ph\text{-}env}} + H_{\mathrm{SR}},
\end{align}
where $H_{\mathrm{dot}}$, $H_{\mathrm{lead}}$, and $H_{\mathrm{lead\text{-}dot}}$ describe the CNT quantum dot, the electronic leads, and the lead-dot tunneling interaction, respectively. $H_{\mathrm{ph}}$ represents the phonons of the torsional mode, which are coupled to the thermal phonon environment $H_{\mathrm{env}}$ via the interaction term $H_{\mathrm{ph\text{-}env}}$. Finally, $H_{\mathrm{SR}}$ denotes the spin-rotation interaction originating from the gyromagnetic effect.
In the subsequent subsections, we present an explicit form of each Hamiltonian.

\subsection{Electron systems}

At cryogenic temperatures, the CNT can be regarded as a quantum dot with discrete electronic levels.
For simplicity, we consider a single-level quantum dot, whose Hamiltonian is given by
\begin{align}
H_{\mathrm{dot}} &= \left(\epsilon_d + \frac{h}{2} \right) d^\dagger_\uparrow d_\uparrow + \left(\epsilon_d - \frac{h}{2}\right) d^\dagger_\downarrow d_\downarrow  + U n_{\uparrow} n_{\downarrow} , 
\end{align}
where \( d_\sigma \) is an annihilation operator for electrons with spin $\sigma$  ($\sigma=\uparrow,\downarrow$), $n_\sigma = d^\dagger_\sigma d_\sigma$ is an electron number operator, \( \epsilon_d \) is the single-particle energy level of the quantum dot, and \( h = g\mu_{\rm B} B \) ($g$: the $g$-factor, $\mu_{\rm B}$: the Bohr magneton) denotes the Zeeman energy induced by the external magnetic field. 

\begin{figure}[t]
\centering
\includegraphics[width=8.5cm]{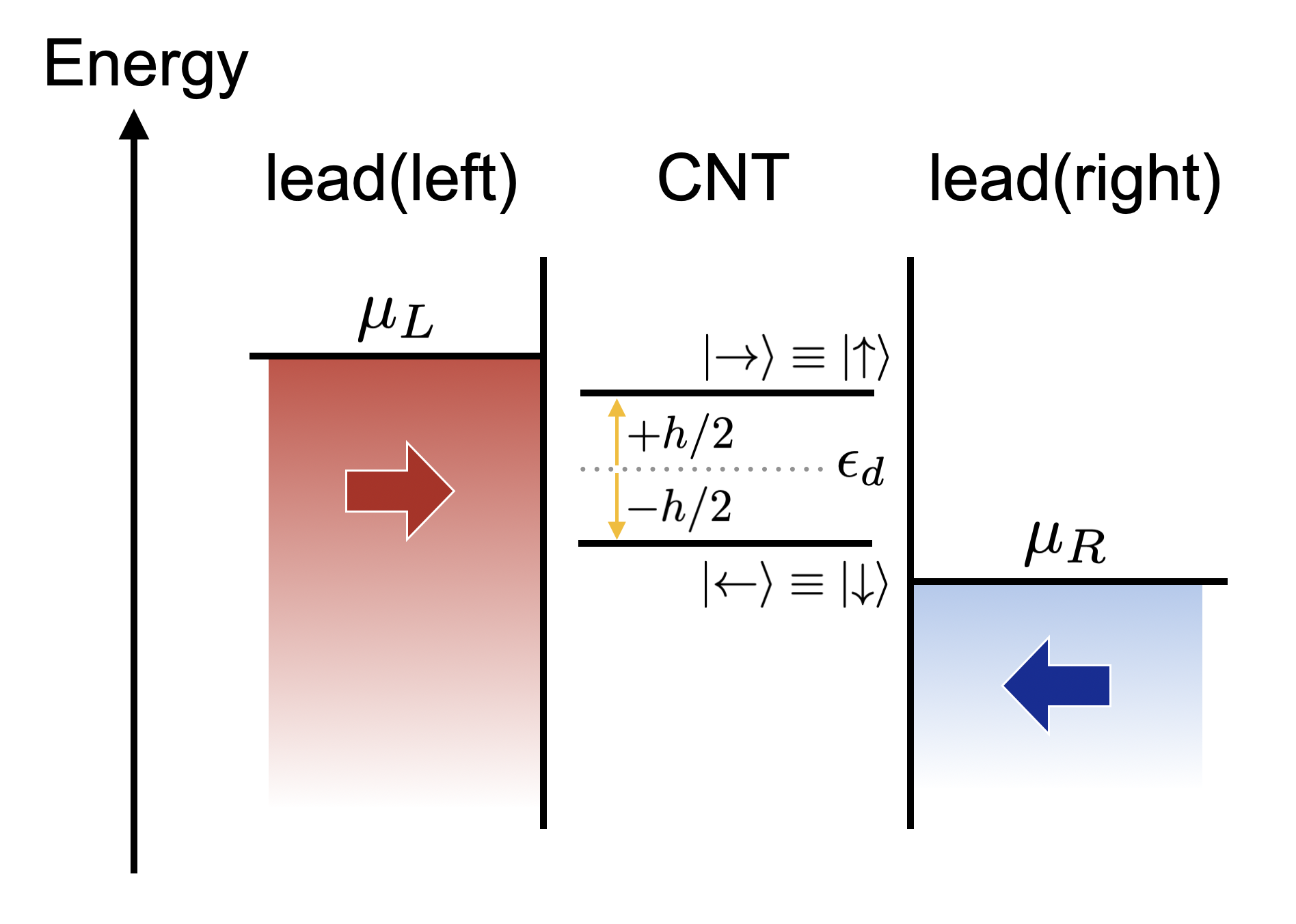}
\caption{Energy-level diagram of the CNT quantum dot. }
\label{fig:energy_diagram}
\end{figure}

We neglect the valley degree of freedom in the CNT in the present minimal model. The conditions under which this simplified description is justified are discussed in Sec.~\ref{sec:soi}.
Following the standard convention in quantum-dot systems, electron spins parallel or antiparallel to the magnetic field are denoted as $\uparrow$ and $\downarrow$, respectively.
The Coulomb interaction $U$ is assumed to be sufficiently large that only the unoccupied state (denoted by $0$) and the one-electron occupied states (denoted by $\uparrow$ and $\downarrow$) are relevant. 
The energies of these states are given by $\epsilon_0 = 0$, $\epsilon_\uparrow = \epsilon_d + h/2$, and $\epsilon_\downarrow = \epsilon_d - h/2$.
The energy level diagram of the quantum dot is shown in Fig.~\ref{fig:energy_diagram}.

The coupling between the CNT and the completely spin-polarized (half-metallic) leads is described by the lead and tunneling Hamiltonian
\begin{align}
H_{\mathrm{lead}} &= \sum_{{\bm k}} (\epsilon_{\bm k}-\mu_{\mathrm{L}}) c_{L{\bm k}\uparrow}^\dagger  c_{L{\bm k}\uparrow} \notag \\
& \hspace{5mm} + \sum_{{\bm k}} 
(\epsilon_{\bm k}-\mu_{\mathrm{R}}) c_{R{\bm k}\downarrow}^\dagger  c_{R{\bm k}\downarrow} , \\
H_{\mathrm{lead\text{-}dot}} &= \sum_{\bm k} \left( \Delta_L d_\uparrow^\dagger c_{L {\bm k}\uparrow}  + \Delta_R d_\downarrow^\dagger c_{R {\bm k}\downarrow}  + {\rm H.c.} \right),
\end{align}
where $c_{L {\bm k}\uparrow}$ and $c_{R {\bm k}\downarrow}$ denote annihilation operators for electrons with wave vector ${\bm k}$ in the two electrodes, $\mu_{\mathrm{L}}$ and $\mu_{\mathrm{R}}$ are the chemical potentials of the left and right leads, and $\Delta_L$ and $\Delta_R$ are tunneling amplitudes.

\subsection{Torsional vibration}

We first consider the torsional (twisting) vibration modes of the CNT within the classical continuum model.
Due to its cylindrical symmetry, the vibration mode is described by a twist angle \( \theta(z',t) \), where a new coordinate \( (x',y',z') \) is introduced so that the \( z' \) axis is aligned with the axis of the CNT.
The restoring torque per unit length resulting from shear deformation is given by
\begin{equation}
\tau(z') = G R^2 S \frac{\partial^2 \theta}{\partial z'^2},
\end{equation}
where \( R \) is the CNT radius, $S=2\pi R d$ is a cross-sectional area of the CNT, $d$ is the wall thickness, and \( G \) is the shear modulus.
Assuming a hollow cylinder and linearized dynamics, the equation of motion reads
\begin{equation}
\frac{\partial^2 \theta}{\partial t^2} = c^2 \frac{\partial^2 \theta}{\partial z'^2}, \qquad c = \sqrt{\frac{G S}{\rho}}, \label{waveequation}
\end{equation}
where \( \rho \) is the mass density of the CNT per unit length.
For a CNT of length $L$, clamped boundary conditions \( \theta(0,t) = \theta(L,t) = 0 \) lead to quantized normal modes of the form
\begin{equation}
\theta(z',t) = \sum_{n=1}^{\infty} \sqrt{\frac{2}{L}} \sin\left( \frac{n\pi z'}{L} \right) \theta_n(t).
\label{theta_quantization}
\end{equation}
In the following, we consider only the fundamental mode ($n=1$) and denote its angular frequency as $\omega_0 = \pi c/L$.

Next, we consider the quantization of the fundamental torsional mode.
The phonon Hamiltonian reads
\begin{equation}
H_{\mathrm{ph}} = \hbar \omega_0 \left( a^\dagger a + \frac{1}{2} \right),
\label{eq:Hph}
\end{equation}
where \( a \) and \( a^\dagger \) are bosonic annihilation and creation operators defined as
\begin{align}
\hat{a} = \sqrt{\frac{\rho R^2 \omega_0}{2\hbar}}\bigg\lbrack\hat{\theta}+i\frac{\hat{\pi}}{\rho R^2 \omega_0}\bigg\rbrack , 
\label{a_operators2}
\\
\hat{a}^\dagger = \sqrt{\frac{\rho R^2 \omega_{0}}{2\hbar}}\bigg\lbrack\hat{\theta}-i\frac{\hat{\pi}}{\rho R^2 \omega_0}\bigg\rbrack .
\label{a_operators}
\end{align}
Here, $\hat{\theta}$ ($= \hat{\theta}_1$) is the operator representing the amplitude of the fundamental torsional mode, and $\hat{\pi}$ is an operator canonically conjugate to $\hat{\theta}$, which satisfies $[\hat{\theta},\hat{\pi}] = i\hbar$.
For a detailed derivation, see the Appendix.
From Eq.~(\ref{eq:Hph}), the time evolution of the twist angle operator $\hat{\theta}(z',t)$ is given by
\begin{align}
\hat{\theta}(z',t) &\simeq \sqrt{\frac{2}{L}} \sin\left( \frac{\pi z'}{L} \right) \hat{\theta}(t), \label{eq:theta1} \\
\hat{\theta}(t) &= \sqrt{\frac{\hbar}{2\rho R^2 \omega_{0}}}(\hat{a} e^{-i\omega_{0} t} +\hat{a}^\dagger e^{i\omega_{0} t}) .
\label{eq:theta2} 
\end{align}

To describe the damping of the torsional mode, we consider energy dissipation into a phonon environment associated with the junctions.
We model this damping with an environment described by the Hamiltonian
\begin{align}
H_\mathrm{env} &= \sum_{\bm q} \hbar \omega_{\bm q} \left( b_{\bm q}^\dagger b_{\bm q} + \frac12 \right) , \\
H_\mathrm{ph\text{-}env} &= \sum_{\bm q} \kappa (a^\dagger b_{\bm q} + a b^\dagger_{\bm q}),
\end{align}
where $b_{\bm q}$ and $b_{\bm q}^\dagger$ are annihilation and creation operators for the bath phonon modes with wave vector $\bm q$, respectively, and $\kappa$ denotes the coupling constant between the torsional mode and the environmental phonon modes.

\subsection{Spin-rotation coupling}

The key mechanism for the conversion between electron spins and torsional vibrations (phonons) is the gyromagnetic effect, which is the conversion of angular momentum between spin and mechanical rotation.
This angular-momentum conversion is expressed by the spin-rotation coupling, $-\bm \omega \cdot \hat{\bm S}$, where $\bm \omega$ is the angular velocity of the rigid-body rotation and $\hat{\bm S}$ is the total spin in the target material.
This spin-rotation coupling can be extended to local mechanical rotation expressed by ${\bm \omega}({\bm r}) = \hat{\bm \Omega}({\bm r})/2$, where $\hat{{\bm \Omega}}({\bm r})$ is a vorticity defined by
\begin{align}
    \hat{\bm{\Omega}}({\bm r}) = \nabla \times \hat{\bm v}({\bm r}) .
\end{align}
Here, ${\bm r}= x' {\bm e}_{x'} + y' {\bm e}_{y'}+z' {\bm e}_{z'}$ is the position on the CNT, and \( \hat{\bm v}({\bm r}) \) is the velocity field defined by
\begin{equation}
\hat{\bm{v}}({\bm r}) = R \left. \frac{\partial \hat{\theta}(z',t)}{\partial t} \right|_{t=0} \, {\bm e}_\perp ,
\end{equation}
with a unit tangent vector \( {\bm e}_\perp \) of the CNT.
Substituting Eqs.~\eqref{eq:theta1} and \eqref{eq:theta2} into this expression, the vorticity is rewritten as
\begin{equation}
\hat{\boldsymbol{\Omega}}({\bm r}) = i\sqrt{\frac{\hbar\omega_0}{L\rho R^2}}(\hat{a} -\hat{a}^\dagger)
\left( \begin{array}{c} 
(\pi x'/L) \cos (\pi z'/L) \\
(\pi y'/L)\cos (\pi z'/L) \\-2\sin(\pi z'/L) \end{array} \right). \label{eq:Omegar}
\end{equation}
The spin-rotation coupling in the CNT is described by the Hamiltonian
\begin{align}
& H_{\mathrm{SR}} = -\frac{R d}{2}\int_0^L \! dz' \int_0^{2\pi}\! d\theta\, \hat{{\bm \Omega}}({\bm r})\cdot\hat{\boldsymbol{S}}(\bm{r}),
\label{H_SR} \\
& \hat{\bm{S}}(\bm{r}) = \frac{\hbar}{2} (\hat{\Psi}^\dagger_\uparrow(\bm{r})\ \hat{\Psi}^\dagger_\downarrow(\bm{r})) \bm \sigma \left( \begin{array}{c} \hat{\Psi}_\uparrow(\bm{r}) \\ \hat{\Psi}_\downarrow(\bm{r}) \end{array} \right),
\end{align}
where $(\theta, z')$ denotes the position on the  CNT in the cylindrical coordinate system, $\hat{\Psi}_\sigma({\bm r})$ is an electron annihilation operator at ${\bm r}$ for spin \( \sigma \),  \( \hat{\bm S}({\bm r}) \) is the electron spin density operator, and $\bm \sigma = ({\sigma}_x, {\sigma}_y, {\sigma}_z)$ represents the Pauli matrices.
By focusing on one eigenstate $\psi({\bm r})$ in the CNT as a quantum-dot level, the electron field operator is approximated as $\hat{\Psi}_\sigma({\bm r}) = \psi({\bm r}) d_\sigma$.
For simplicity, we neglect the valley degree of freedom and assume a uniform wavefunction $\psi({\bm r})=1/\sqrt{2\pi R dL}$.
After performing the integral given in Eq.~\eqref{H_SR} with Eq.~\eqref{eq:Omegar}, the Hamiltonian $H_{\rm SR}$ is given by
\begin{equation}
H_\mathrm{SR} =  \frac{2i}{\pi R}\sqrt{\frac{\hbar \omega_0}{L\rho}} (\hat{a} -\hat{a}^\dagger) \hat{s}_{z'},
\label{H_SR_tilt0}
\end{equation}
where $\hat{s}_{z'}$ is the $z'$-component of the electron spin in the quantum dot ($\hat s_{z'}=\hat{\boldsymbol s}\cdot\boldsymbol e_{z'}$).
The spin-rotation coupling remains only in the $z'$-direction, which corresponds to the axis of the CNT.
This means that an effective ac magnetic field induced by the CNT torsional modes appears only in the $z'$ direction.
The spin-rotation coupling is rewritten with the spin components in the laboratory coordinates ($\hat{\bm s}=\hat{s}_{x} {\bm e}_{x}+\hat{s}_{y} {\bm e}_{y}+\hat{s}_{z} {\bm e}_{z}$) as
\begin{equation}
H_\mathrm{SR} =  \frac{2i}{\pi R} \sqrt{\frac{\hbar \omega_0}{L\rho}} (\hat{a} -\hat{a}^\dagger) 
\left[ \cos\varphi  \hat{s}_z + \frac{\sin\varphi}{2} (\hat{s}_+ + \hat{s}_-) \right],
\label{H_SR_tilt}
\end{equation}
where $\varphi$ is the tilting angle  (see Fig.~\ref{fig:model}) and the spin operators are related to the creation and annihilation operators of electrons in the quantum dot as $\hat{s}_z = (\hbar/2) (d_\uparrow^\dagger d_\uparrow - d_\downarrow^\dagger d_\downarrow)$, $\hat{s}_+ = \hbar d_\uparrow^\dagger d_\downarrow$, and $\hat{s}_- = (\hat{s}_+)^\dagger$.
The tilting of the CNT axis with respect to the magnetization of the electrodes is required to produce the spin-flipping terms in $H_\mathrm{SR}$, which enable mutual conversion between the spin angular momentum of electrons and the mechanical angular momentum of the CNT.

\section{Formulation}\label{sec:method}

In this work, we calculate transport properties by the method of the master equation.
The fundamental torsional mode is regarded as a harmonic oscillator, and its quantum state is denoted with $m$ ($=0,1,2,\cdots$).
Combining this with the electronic state in the quantum dot, we label the eigenstates as $(0,m)$, $(\uparrow,m)$, and $(\downarrow,m)$.
In the following, we calculate the transition rates between the states based on Fermi's golden rule using perturbative treatment with respect to $H_\mathrm{ph\text{-}env}$, $H_{\mathrm{lead\text{-}dot}}$, and $H_{\mathrm{SR}}$.

\subsection{Phonon energy relaxation}

The relaxation and excitation by the phonon bath are expressed by the transition process induced by $H_\mathrm{ph\text{-}env}$.
The relaxation rate from $(\alpha,m)$ to $(\alpha,m-1)$, keeping the electronic state $\alpha$ ($=0,\uparrow,\downarrow$) unchanged, is calculated as
\begin{align}
& \Gamma_{(\alpha,m)\rightarrow (\alpha,m-1)}
= \frac{2\pi}{\hbar^2}\kappa^2 m \mathcal{D}(\omega_0)
\left[1+f_{\mathrm{BE}}(\omega_0)\right],
\end{align}
where $\mathcal{D}(\omega_0)$ denotes the phonon density of states of the bath evaluated at the system frequency $\omega_0$, $f_{\mathrm{BE}}(\omega) = (e^{\hbar \omega/k_{\rm B}T}-1)^{-1}$ is the Bose--Einstein distribution function, $k_{\rm B}$ is the Boltzmann constant, and we have assumed that the phonon bath is in thermal equilibrium at temperature $T$. 
Similarly, the excitation rate  from $|\alpha,m-1\rangle$ to $|\alpha,m\rangle$ is calculated as
\begin{align}
\Gamma_{(\alpha,m-1)\rightarrow (\alpha,m)}
= \frac{2\pi}{\hbar^2}\kappa^2 m \mathcal{D}(\omega_0)
f_{\mathrm{BE}}(\omega_0).
\end{align}
In the following, we denote the prefactor as
\begin{align}
k_{\mathrm{relax}} \equiv \frac{2\pi}{\hbar^2}\kappa^2 \mathcal{D}(\omega_0).
\end{align}
We assume that $k_{\rm relax}$ is proportional to $\omega_0$, corresponding to an Ohmic bath.
At sufficiently low temperatures ($k_{\rm B} T \ll \hbar \omega_0$), $k_{\mathrm{relax}}$ corresponds to the mechanical damping rate $\gamma_{\mathrm{mech}} = \omega_0/Q$, where $Q$ is the quality factor of the torsional mode.
The realistic value of $Q$ is expected to range from $10^{2}$ to $10^{5}$ in experiments at low temperatures, based on experiments on torsional and flexural modes~\cite{Hall2008, Huttel2009}.

\subsection{Electron transfer via lead-dot coupling}

The electron transfer rates between the CNT and the half-metallic electrodes are calculated from the lead-dot interaction $H_{\mathrm{lead\text{-}dot}}$.
The transition rate for transferring electrons from the left lead to the empty dot is calculated as 
\begin{align}
\Gamma_{(0,m) \rightarrow (\uparrow,m)}
= \Gamma_{\mathrm{L}}\, f_{\mathrm{L}}\!\left(\epsilon_d + \frac{h}{2}\right),
\label{eq:Gamma1}
\end{align}
where $f_{\mathrm{L}}(\epsilon) = (e^{(\epsilon-\mu_{\mathrm{L}})/k_{\rm B}T}+1)^{-1}$ is the Fermi--Dirac distribution of the left lead, $\Gamma_{\mathrm{L}} = 2\pi \Delta_L^2 D(\epsilon_{\rm F})/\hbar$ denotes the tunneling rate, and $D(\epsilon_{\rm F})$ is the density of states at the Fermi energy.
Here, we have replaced the momentum sum with an energy integral as
$\sum_{{\bm k}} (\cdots) \simeq D(\epsilon_{\rm F}) \int d\epsilon\, (\cdots)$. 
Similarly, the reverse process, in which a spin-$\uparrow$ electron transfers from the dot into the left lead, occurs at the rate
\begin{align}
\Gamma_{(\uparrow,m) \rightarrow (0,m)}
= \Gamma_{\mathrm{L}} \left[ 1 - f_{\mathrm{L}}\!\left(\epsilon_d + \frac{h}{2}\right) \right].
\label{eq:Gamma2}
\end{align}
The tunneling rates involving the right lead and the spin-$\downarrow$ state
are obtained in a manner analogous to that of the left lead.
\begin{align}
\Gamma_{(0,m) \rightarrow (\downarrow,m)}
&= \Gamma_{\mathrm{R}}\, f_{\mathrm{R}}\!\left(\epsilon_d - \frac{h}{2}\right),\label{eq:Gamma3} \\
\Gamma_{(\downarrow,m) \rightarrow (0,m)} 
&= \Gamma_{\mathrm{R}} \left[ 1 - f_{\mathrm{R}}\!\left(\epsilon_d - \frac{h}{2}\right) \right],\label{eq:Gamma4}
\end{align}
where $f_{\mathrm{R}}(\epsilon) = (e^{(\epsilon-\mu_{\mathrm{R}})/k_{\rm B}T}+1)^{-1}$.

\subsection{Spin-phonon conversion at a dot}

We note that there is no current through a dot solely due to electron transfer from the lead-dot coupling.
To obtain a finite current, a spin-flip process at the dot is required.
The spin-flip process due to the spin-rotation coupling $H_{\rm SR}$ involves phonon absorptions and emissions.
Based on Fermi's golden rule, the transition rate from $(\uparrow,m)$ to $(\downarrow,m+1)$ is calculated as
\begin{align} 
\Gamma_{(\uparrow, m) \rightarrow (\downarrow, m+1)} &=  \Gamma_{\rm{SR}}^+(m+1), \label{Gamma_SR_plus}\\
\Gamma_{(\downarrow, m+1) \rightarrow (\uparrow, m)} &=  \Gamma_{\rm{SR}}^-(m+1), 
\label{Gamma_SR_minus}
\end{align} 
where $\Gamma^\pm_{\mathrm{SR}} = A \delta(h- \hbar\omega_0)$ and $A$ is given by 
\begin{align} 
A &= \frac{2\hbar^2 \omega_0}{\pi R^2 L \rho} \sin^2 \varphi \notag \\
&= \frac{2\hbar^2 \omega_0^2}{\pi^2 R^2}\sqrt{\frac{1}{GS\rho}}\sin^2 \varphi .
\label{def_A}
\end{align}
For actual calculations, we replace the delta function with a Lorentzian function as~\cite{Schoeller1994,Timm2008,BruusFlensberg2016}
\begin{align}
\delta(h - \hbar\omega_0)\rightarrow \frac{\hbar \Gamma/2\pi}{(h-\hbar\omega_0)^2+(\hbar \Gamma/2)^2},
\end{align}
with a broadening factor $\Gamma$, which is induced by both energy relaxation and electron transfer between the leads and the dot.
In the following calculation, we always set the parameters to satisfy $\Gamma_{\mathrm{L}}, \Gamma_{\mathrm{R}} \gg k_{\rm relax}$, for which the choice of $\Gamma = \Gamma_{\mathrm{L}} + \Gamma_{\mathrm{R}}$ is suitable.
To characterize the strength of the spin-rotation coupling, we define the spin-flipping rate on the resonance as
\begin{align}
\Gamma_{\rm SR} \equiv \left. \Gamma_{\rm SR}^\pm \right|_{h= \hbar\omega_0} = 
\frac{4\hbar \omega_0^2}{\pi^3 \Gamma R^2}\sqrt{\frac{1}{GS\rho}}\sin^2 \varphi .
\end{align}
The spin-rotation coupling is enhanced for a small broadening factor $\Gamma$, a small radius $R$, and a large resonant frequency $\omega_0$.

\subsection{Master equation}

To evaluate transport properties, we derive a master equation for the probabilities \( P_{\alpha,m} \) of finding the system in a joint electron-phonon state \( (\alpha,m) \), incorporating the transition rates derived above. 
The master equation is given as
\begin{align}
\dot{P}_{0,m} &= -\Gamma_{\mathrm{L}} f_{\mathrm{L}} P_{0,m} + \Gamma_{\mathrm{L}} (1-f_{\mathrm{L}}) P_{\uparrow,m} \notag \\ 
&+ \Gamma_{\mathrm{R}} (1-f_{\mathrm{R}}) P_{\downarrow,m} - \Gamma_{\mathrm{R}} f_{\mathrm{R}} P_{0,m} + {\cal D}[P_{0,m}], \\
\dot{P}_{\uparrow,m} &= - \Gamma_{\rm SR}^+(m+1) P_{\uparrow,m}+ \Gamma^-_{\rm SR} (m+1) P_{\downarrow,m+1} \notag \\
& + \Gamma_{\mathrm{L}} f_{\mathrm{L}} P_{0,m} - \Gamma_{\mathrm{L}} (1-f_{\mathrm{L}}) P_{\uparrow,m} + {\cal D}[P_{\uparrow,m}], \\ 
\dot{P}_{\downarrow,m} &=- \Gamma_{\rm SR}^- m P_{\downarrow,m}  + \Gamma_{\rm SR}^+ m P_{\uparrow,m-1} \notag \\
& - \Gamma_{\mathrm{R}} (1-f_{\mathrm{R}}) P_{\downarrow,m} + \Gamma_{\mathrm{R}} f_{\mathrm{R}} P_{0,m} + {\cal D}[P_{\downarrow,m}], \\
{\cal D}[P_{\alpha,m}] &= k_{\mathrm{relax}} [ m f_{\mathrm{BE}}P_{\alpha,m-1} -m(1+f_{\mathrm{BE}})P_{\alpha,m}  \notag\\ 
&+ (m+1) (1+f_{\mathrm{BE}}) P_{\alpha,m+1} - (m+1) f_{\mathrm{BE}} P_{\alpha,m}].
\end{align}
The steady-state probability distribution is obtained by setting $\dot{P}_{0,m} = \dot{P}_{\uparrow,m}=\dot{P}_{\downarrow,m}=0$ under the condition
\( \sum_{\alpha,m} P_{\alpha,m} = 1 \).
In the actual calculation, we take a large cutoff $m_{\rm max}$ and consider the probability distribution $P_{\alpha,m}$ in a finite range of $0\le m \le m_{\rm max}$.

The steady-state current from the dot to the left lead is calculated from the probability distribution as
\begin{align}
I = e \Gamma_{\mathrm{L}} \sum_{m}[f_{\mathrm{L}} P_{0,m} - (1-f_{\mathrm{L}}) P_{\uparrow,m}],
\end{align}
where $e$ ($>0$) is the elementary charge.

\subsection{Phonon driving}

In our setup, when a source-drain bias is applied across the two half-metallic electrodes in an antiparallel configuration, only a spin-$\uparrow$ electron can enter the quantum dot from the left electrode, while a spin-$\downarrow$ electron can leave it into the right electrode.
In the absence of spin flips in the quantum dot, no current flows due to the spin valve effect.
Spin-flip transitions due to the SRC become effective when the resonance condition \( \hbar \omega_0 \approx h \) is satisfied, enabling a resonant mechanism for spin-phonon conversion.
This phonon driving can be detected in the charge current through the CNT.

To see the feasibility of the proposed scheme for driving the torsional vibration of the CNT, we roughly estimate the experimental conditions.
We consider a single-wall CNT with a radius \( R \approx 0.5\,\mathrm{nm} \), a wall thickness \( d \approx 0.34\,\mathrm{nm} \), and a length \(L = 100\,\mathrm{nm} \).
The cross-sectional area \( S = 2\pi R d \approx 1.1 \times 10^{-18}\,\mathrm{m}^2 \) and the linear mass density \( \rho \approx 2.4 \times 10^{-15}\,\mathrm{kg}/\mathrm{m} \) are calculated from the surface density of graphene $7.6 \times 10^{-7} \text{ kg/m}^2$. The shear modulus $G$ is taken as $0.41\, \text{TPa}$, based on theoretical calculations and experimental reports~\cite{Lu1997,Hall2016}.
Substituting into the dispersion relation for torsional waves, we find the lowest eigenfrequency of the CNT to be \( \omega_0 \approx 4.3 \times 10^{11}\,\mathrm{rad}/\mathrm{s} \). This frequency corresponds to the phonon energy on the order of \( \hbar\omega_0 \sim 280\,\mu\mathrm{eV} \), which is comparable to the Zeeman splitting at magnetic fields of \( 2.4 \, {\rm T}\).
By choosing \( \hbar\Gamma_{\mathrm{L}} = \hbar\Gamma_{\mathrm{R}} = 2.0\, \mu\mathrm{eV} \), the spin-flip rate at the resonance condition $h = \hbar \omega_0$ is roughly estimated as \( \hbar\Gamma_{\rm SR}\simeq 1.7\times 10^{-2}\, \mu\mathrm{eV} \), which is comparable to the phonon relaxation rate $\hbar k_{\rm relax}$.
This estimate indicates that driving the torsional vibration of the CNT is experimentally feasible under realistic conditions, utilizing efficient spin-phonon conversion at the resonance condition $h=\hbar \omega_0$.

\section{Results}\label{sec:results}

In this section, we present numerical results for the steady-state probability distribution and the corresponding current. 
For simplicity, we set the center of the energy levels as $\epsilon_d = 0$ in the following.
Throughout this section, the shear modulus, the radius of the CNT, the tilting angle, and the tunneling rates are taken as $G=0.41\, \text{TPa}$, $R=0.5 \, \mathrm{nm}$, $\varphi = \pi/4$, and \( \hbar\Gamma_{\mathrm{L}} = \hbar\Gamma_{\mathrm{R}} = 2.0\ \mu\mathrm{eV} \), respectively.

\subsection{Resonant behavior}

\begin{figure}[t]
\centering
\includegraphics[width=8.5cm]{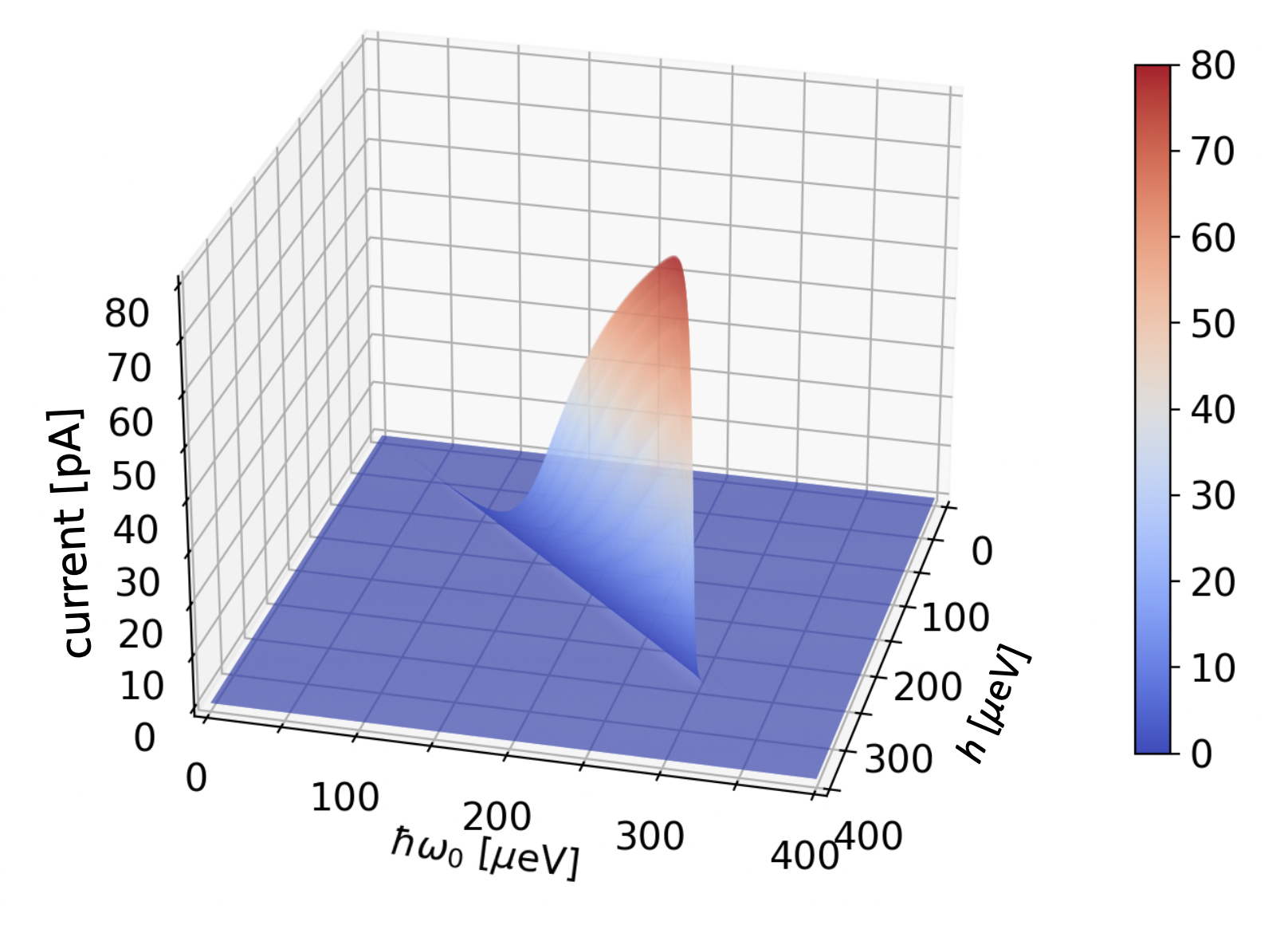}
\caption{Contour plot of the current as a function of $\hbar\omega_0$ and $h$. 
The parameters are set as \( k_BT=2.0\,\mu{\rm eV}\;(T=23.2\,{\rm mK}) \), \( \mu_{\mathrm{L}} = -\mu_{\mathrm{R}} = 150\, \mu\mathrm{eV} \), and $k_\mathrm{relax}/\omega_0 = 1/Q = 3.0\times10^{-5}$.}
\label{fig:3d_current}
\end{figure}

Figure~\ref{fig:3d_current} shows the steady-state current as a function of the Zeeman splitting energy $h$ and the phonon energy $\hbar\omega_0$, where the phonon frequency $\omega_0$ is controlled by the CNT length $L$.
We set the chemical potentials as $\mu_{\mathrm{L}} = -\mu_{\mathrm{R}} = 150\,\mu\mathrm{eV}$, for which the energy levels of the CNT, $\epsilon_\uparrow = h/2$ and $\epsilon_\downarrow = -h/2$, are in the energy window $[\mu_{\mathrm{R}},\mu_{\mathrm{L}}]$, allowing electron transfer between the dot and the two leads when $h< 300\,\mu\mathrm{eV}$.
The other parameters are given in the caption of Fig.~\ref{fig:3d_current}.
The most prominent feature in the figure is a sharp ridge appearing along the diagonal line \( h = \hbar\omega_0 \), on which the energy difference between the spin-$\uparrow$ and spin-$\downarrow$ states matches the phonon energy. 
Under this resonant condition, the spin-rotation coupling efficiently facilitates spin-flip transitions accompanied by phonon emission, bridging the two spin states and allowing a continuous flow of electrons from the left lead to the right lead. The current reaches a maximum of approximately \( 80\,\mathrm{pA} \), which is sufficiently detectable in experiments.

\subsection{Phonon populations}

\begin{figure}[t]
\centering
\includegraphics[width=8.5cm]{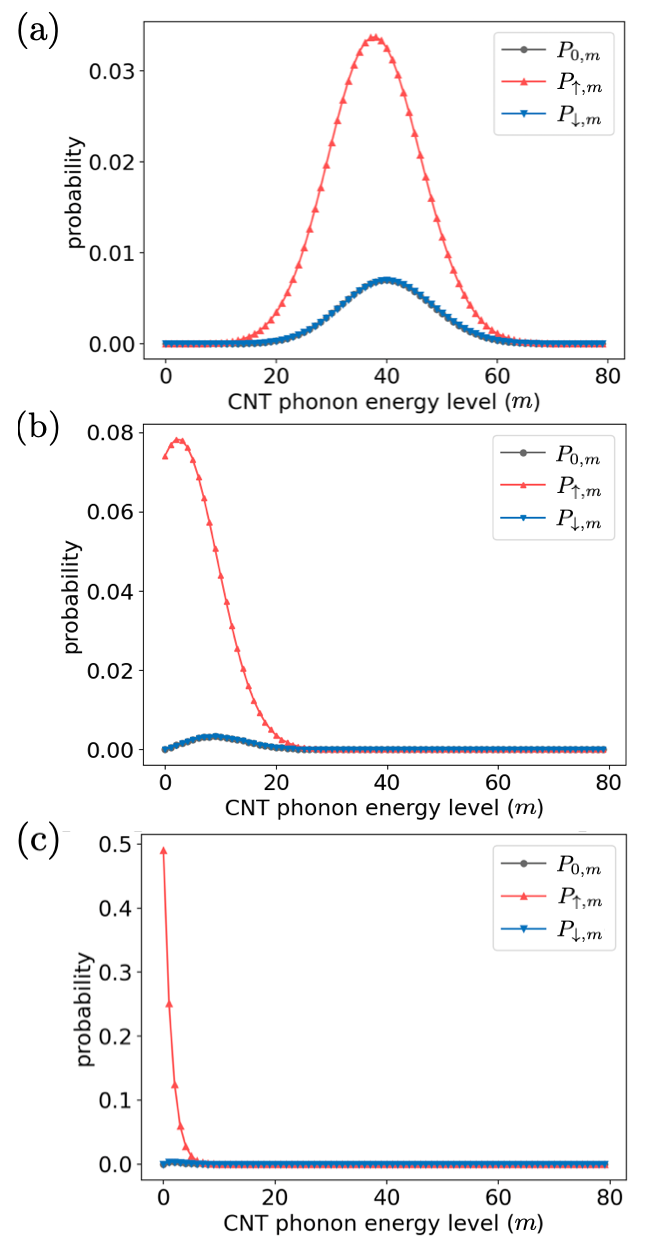}
\caption{Probability distributions of the CNT torsional mode at (a) \( k_\mathrm{relax}/\omega_0 = 1/Q = 3.0\times10^{-5}\), (b) \( k_\mathrm{relax}/\omega_0 = 5.0\times10^{-5}\) and (c) \( k_\mathrm{relax}/\omega_0 = 1.0\times10^{-4}\).
Note that the curves for $P_{0,m}$ and $P_{\downarrow,m}$ almost overlap.
The other parameters are set as \( h=\hbar\omega_0 = 250\, \mu\mathrm{eV} \), \( k_BT=2.0\,\mu{\rm eV}\;(T=23.2\,{\rm mK})\), and  \( \mu_{\mathrm{L}} = -\mu_{\mathrm{R}} = 150\, \mu\mathrm{eV} \).
The torsional-angle amplitude is estimated from the mean phonon occupation $\langle m\rangle$ using Eq.~(\ref{theta_estimate}).}
\label{fig:2d_phonon_dist_T20}
\end{figure}

Figure~\ref{fig:2d_phonon_dist_T20} shows the probability distributions $P_{\alpha,m}$ under the resonance condition $h=\hbar \omega_0$. 
One finds that $P_{\uparrow,m}$ shows an excess over $P_{0,m}$ and $P_{\downarrow,m}$, indicating the successful driving of the torsional mode of the CNT.
The average phonon population depends on the ratio between the spin-flip rate and the phonon relaxation rate because the steady state is achieved by a balance between these two processes.
We will discuss this feature in detail in Sec.~\ref{sec:other_dep}.
The phonon populations show the relation 
$P_{\uparrow,m} > P_{0,m} \simeq P_{\downarrow,m}$, reflecting that the bottleneck of electron transfer through the dot arises from spin flipping due to spin-rotation coupling.

Using this result, we estimate the order of magnitude of the torsional displacement amplitude in the CNT. 
According to Eqs.~\eqref{a_operators2} and \eqref{a_operators}, the amplitude operator for the fundamental torsional mode, $\hat{\theta}$, is given by
\begin{align}
\hat{\theta} = \sqrt{\frac{\hbar}{2\rho R^2 \omega_{0}}}(\hat{a} +\hat{a}^\dagger).
\end{align}
Based on the quantization scheme given in Eq.~\eqref{eq:theta1}, the amplitude of the torsional angle at the center of the CNT is estimated as
\begin{align}
\mathcal{O}(\theta) &\approx  \sqrt{\frac{2}{L}\langle\hat{\theta}^2\rangle} 
= \frac{1}{R}\sqrt{\frac{\hbar}{\pi}\sqrt{\frac{1}{\rho GS}}(2\langle m\rangle+1)}.
\label{theta_estimate}
\end{align}
For $k_{\mathrm{relax}}/\omega_0 = 1/Q = 3.0\times10^{-5}$, $5.0\times10^{-5}$, and $1.0\times10^{-4}$, 
we find $\langle m\rangle \approx 40$, $7$, and $1$, respectively. 
These values correspond to $\mathcal{O}(\theta) \approx 1.1$, $0.5$, and $0.2$ degrees.

\subsection{Bias voltage dependence}

\begin{figure}[t]
\centering
\includegraphics[width=8.5cm]{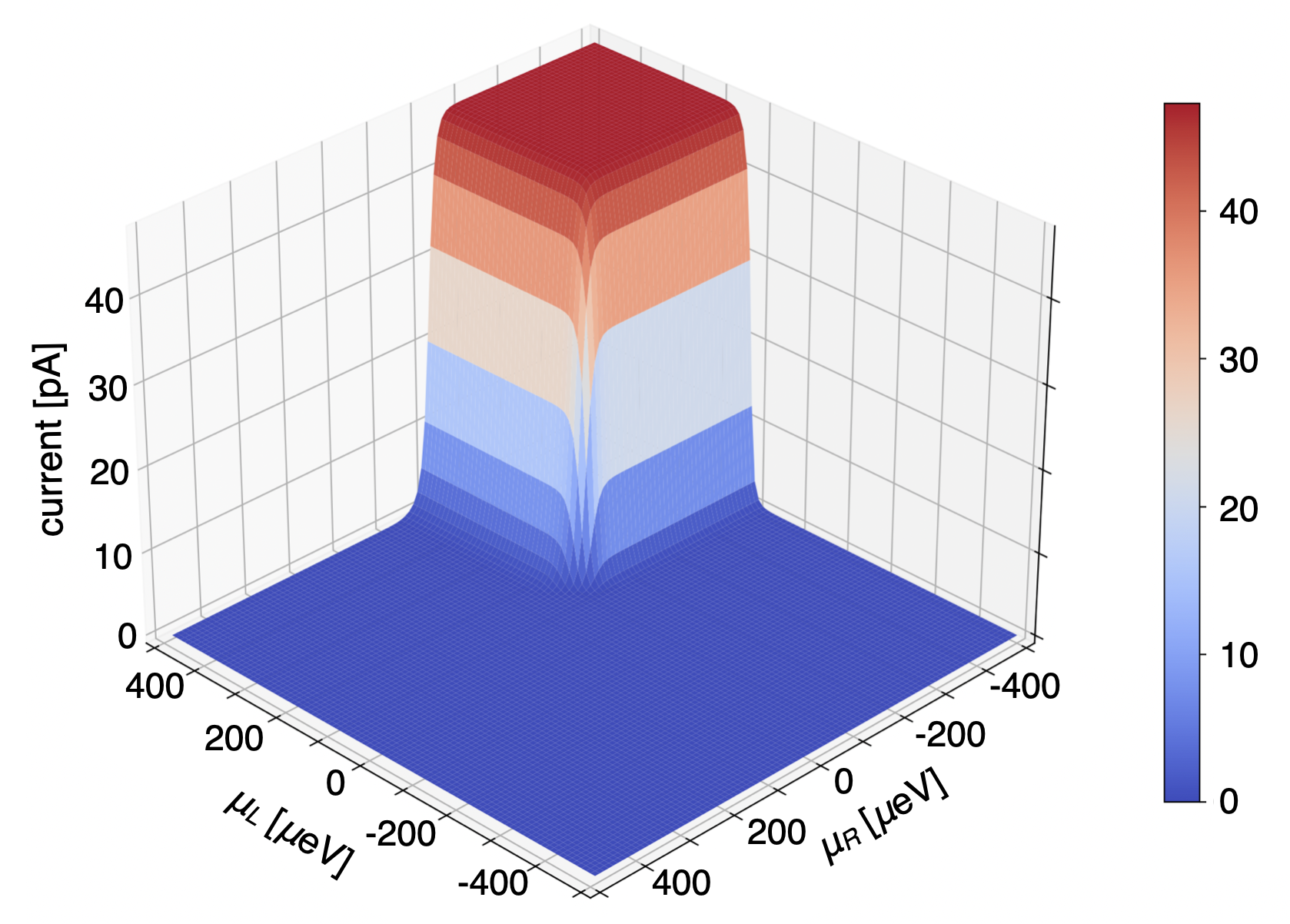}
\caption{Contour plot of the current as a function of $\mu_{\mathrm{L}}$ and $\mu_{\mathrm{R}}$.
The parameters are set as \( h = \hbar\omega_0 = 200\, \mu\mathrm{eV} \), \( k_BT=10.0\,\mu{\rm eV}\;(T=116\,{\rm mK}) \), and
\( k_\mathrm{relax}/\omega_0 = 3.0\times10^{-5}\).}
\label{fig:3d_current_muLR}
\end{figure}

Figure~\ref{fig:3d_current_muLR} shows a contour plot of the current as a function of $\mu_{\mathrm{L}}$ and $\mu_{\mathrm{R}}$ for $h=\hbar \omega_0$.
As already mentioned, the current flows when the energies of the two quantum levels, $\epsilon_{\uparrow} = h/2$ and $\epsilon_{\downarrow} = -h/2$, are within the bias window $[\mu_{\mathrm{R}},\mu_{\mathrm{L}}]$
(see also Fig.~\ref{fig:energy_diagram}).
This expectation can be verified by Fig.~\ref{fig:3d_current_muLR}, which shows that the current flows when both $\mu_{\mathrm{L}} > h/2$ and $\mu_{\mathrm{R}} < -h/2$ are satisfied.
Once this condition is well fulfilled, the current  saturates around \(50 \, \mathrm{pA}\) in the present parameter set.
The transition width from the zero-current state to the current saturation is determined by the temperature $k_{\mathrm{B}} T$.

\subsection{Other parameter dependences}
\label{sec:other_dep}

\begin{figure}[t]
\centering
\includegraphics[width=8.5cm]{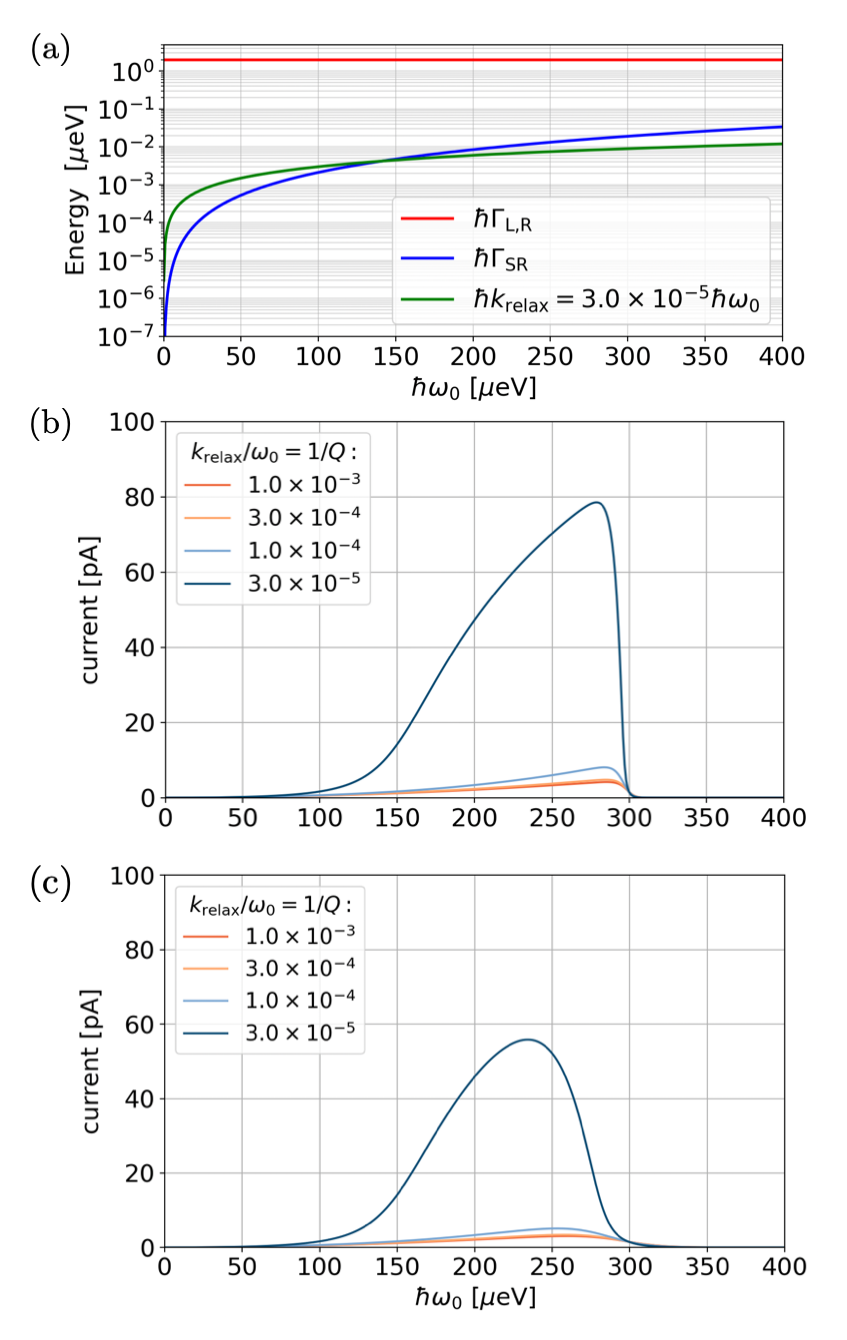}
\caption{(a) Comparison of the energy scales for $\hbar\Gamma_{\mathrm{L},\mathrm{R}}$, $\hbar\Gamma_{\text{SR}}$ and $\hbar k_{\text{relax}}$ as a function of  $\hbar\omega_0$. The damping rate is set as $k_\mathrm{relax}/\omega_0 = 1/Q =  3.0\times10^{-5}$.
Current under the resonant condition ($h=\hbar \omega_0$) is plotted as a function of $\hbar\omega_0$ at (b) \( k_BT=2.0\,\mu{\rm eV}\;(T=23.2\,{\rm mK}) \) and (c) \(k_BT=10.0\,\mu{\rm eV}\;(T=116\,{\rm mK}) \). The other parameters are set as \( \mu_{\mathrm{L}} = -\mu_{\mathrm{R}} = 150.0\, \mu \mathrm{eV} \).}
\label{fig:2d_current_krelax_T02_10}
\end{figure}

Figure~\ref{fig:2d_current_krelax_T02_10}(a) shows the tunneling rate $\Gamma_{\mathrm{L,R}}$ ($=\Gamma_{\mathrm{L}}=\Gamma_{\mathrm{R}}$), the relaxation rate $k_{\rm relax}$, and the spin-phonon conversion rate $\Gamma_{\rm SR}$ as functions of the phonon energy $\hbar\omega_0$.
Within the plotted range of $\hbar\omega_0$, $\Gamma_{\mathrm{L,R}}$ is always much larger than $\Gamma_{\rm SR}$ and $k_{\rm relax}$.
At low phonon frequencies, the phonon damping $k_{\rm relax}$ exceeds the spin-flip rate $\Gamma_{\rm SR}$, leading to a strong suppression of phonon driving.
On the other hand, $\Gamma_{\rm SR}$ becomes larger than $k_{\rm relax}$ for $\hbar\omega_0 > 140\,\mu\mathrm{eV}$, allowing efficient phonon driving.

Figure~\ref{fig:2d_current_krelax_T02_10}(b) shows the current as a function of $\hbar\omega_0$ at $k_BT=2.0\,\mu{\rm eV}\;(T=23.2\,{\rm mK})$ under the resonant condition $h=\hbar\omega_0$. 
The other parameters are given in the caption of Fig.~\ref{fig:2d_current_krelax_T02_10}.
The four curves correspond to $k_{\rm relax}/\omega_0 = 3\times10^{-5}$, $1\times10^{-4}$, $3\times10^{-4}$, and $1\times10^{-3}$, respectively.
We note that $k_{\rm relax}/\omega_0$ is approximately the inverse of the quality factor of the CNT torsional mode.
In the low-frequency region ($\hbar\omega_0 < 140\,\mu\mathrm{eV}$), the current is suppressed by strong phonon damping (see also Fig.~\ref{fig:2d_current_krelax_T02_10}(a)).
In the moderate-frequency region ($140\,\mu\mathrm{eV} < \hbar\omega_0 < 300\,\mu\mathrm{eV}$), spin-flip processes due to the SRC exceed the phonon damping, leading to a remarkable increase in the current.
We note that $\Gamma_{\rm SR}$ is proportional to $\omega_0^2$ according to Eq.~(\ref{def_A}).
In the high-frequency region ($\hbar\omega_0 > 300\,\mu\mathrm{eV}$), the current drops again because the energies of the dot levels go outside the bias window $[\mu_{\mathrm R},\mu_{\mathrm L}]$.

Figure~\ref{fig:2d_current_krelax_T02_10}(c) shows the frequency dependence of the current at a higher temperature $k_BT=10.0\,\mu{\rm eV}\;(T=116\,{\rm mK})$.
Although the overall features are similar to the low-temperature result in Fig.~\ref{fig:2d_current_krelax_T02_10}(b), the current is reduced.
This reduction originates from the thermal broadening of the Fermi--Dirac distribution in the electrodes, as indicated by Eqs.~\eqref{eq:Gamma1}--\eqref{eq:Gamma4}.

\section{Discussion}
\label{sec:discussion}

\subsection{Realistic ferromagnetic electrodes}
\label{sec:partial}

\begin{figure}[tb]
\centering
\includegraphics[width=8.5cm]{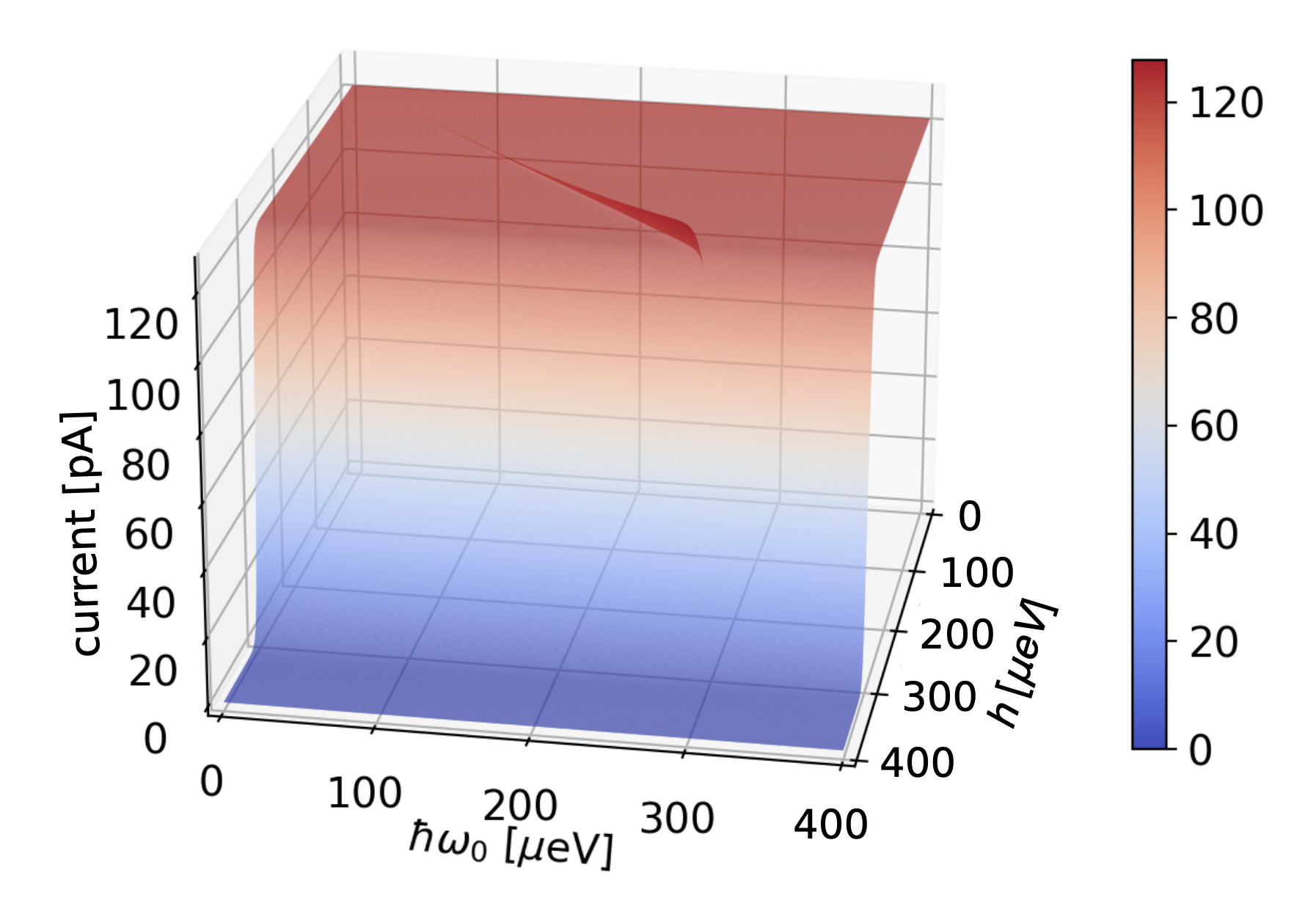}
\caption{Contour plot of the current as a function of $\hbar\omega_0$ and $h$ for a finite spin polarization $P = 0.45$. The parameters are set as \( \hbar\bar{\Gamma}_{\mathrm{L}} = \hbar\bar{\Gamma}_{\mathrm{R}} = 1.0\ \mu\mathrm{eV} \) and all other parameters are the same as in Fig.~\ref{fig:3d_current}.}
\label{fig:3d_current_polarized}
\end{figure}

Throughout the preceding sections, we have assumed half-metallic electrodes with perfect spin polarization for simplicity.
For experimental realization, however, it is more realistic to use ferromagnetic metals with partial spin polarization, such as Fe, Co, Ni, Permalloy, or related alloys.
Ferromagnetic contacts to CNTs have been realized experimentally in a number of spin-transport devices, including CNT spin valves and ferromagnetically contacted CNT quantum dots~\cite{Tsukagoshi1999,Sahoo2005,Jensen2005,Hauptmann2008,Aurich2010,Samm2014}.
These experiments indicate that the use of ferromagnetic electrodes in CNT transport devices is experimentally well established.

Our formulation can be extended to such partially polarized electrodes by modifying the transition rates between the electrodes and the quantum dot as
\begin{align}
\Gamma_{(0,m)\rightarrow(\sigma,m)} &= \bar{\Gamma}_{\mathrm{L}} (1+\sigma P) f_{\mathrm{L}}\left( \epsilon_d + \sigma \frac{h}{2} \right) \notag \\
& + \bar{\Gamma}_{\mathrm{R}} (1-\sigma P) f_{\mathrm{R}}\left( \epsilon_d + \sigma \frac{h}{2} \right), 
\end{align}
where $P$ is the spin polarization of the electrodes at the Fermi level, $\sigma = +1(-1)$ corresponds to spin $\uparrow(\downarrow)$ and $\bar{\Gamma}_{\mathrm{L},\mathrm{R}}$ are the averaged transition rates per spin.
The transition rates for the inverse processes are obtained by replacing $f_{\mathrm{R},\mathrm{L}}$ with $1-f_{\mathrm{R},\mathrm{L}}$.
As a representative value for realistic ferromagnetic electrodes, we use $P=0.45$, which is typical of the transport spin polarization reported for common ferromagnetic metals~\cite{Soulen1998}.

Figure~\ref{fig:3d_current_polarized} shows the current as a function of the Zeeman splitting energy $h$ and the phonon energy $\hbar \omega_0$ for $P=0.45$.
In contrast to the half-metallic case, the current remains finite even away from the resonance because the finite minority-spin density of states in the electrodes allows spin-conserving leakage channels.
Nevertheless, the SRC-induced resonant current enhancement along $h=\hbar\omega_0$ remains clearly visible on top of this background.
This result indicates that the proposed resonance should remain experimentally distinguishable even for partially spin-polarized ferromagnetic contacts.

A similar off-resonant background may also arise if the spin quantization axis of the relevant CNT eigenstates is not collinear with the polarization axes of the ferromagnetic electrodes. 
In this case, the dot eigenstates can have finite overlap with both lead spin channels, providing a leakage channel that does not require SRC-induced spin flips. 
Such a background contribution is not tied to the torsional eigenfrequency, whereas the SRC-induced contribution appears as a resonant enhancement at $h=\hbar\omega_0$ and depends on the mechanical damping rate. 
Thus, as in the case of partially polarized ferromagnets, the resonant enhancement provides a way to distinguish the SRC-induced current from a nonresonant background.

\subsection{Spin-orbit interaction}
\label{sec:soi}

\begin{figure}[t]
\centering
\includegraphics[width=8.5cm]{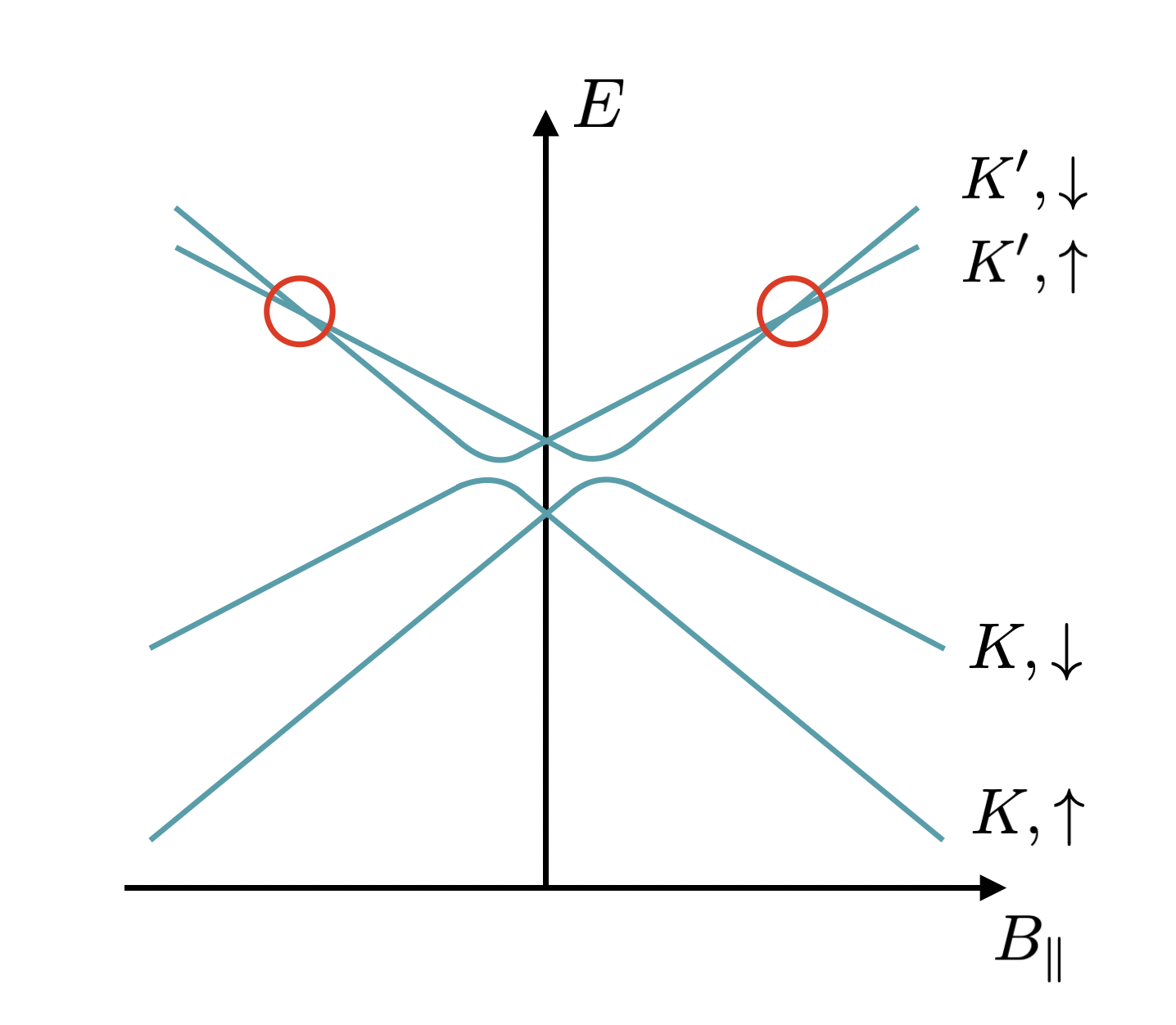}
\caption{Schematic energy-level diagram of a CNT quantum dot including valley degrees of freedom and SOI. 
The red circles indicate the magnetic-field regime in which an isolated pair of spin states within a single valley dominates transport and can be described by the effective two-level model used in this work.
}
\label{fig:magnetic_region}
\end{figure}

In the present work, we employ a simplified single-level model for the CNT quantum dot. 
In realistic CNT quantum dots, however, the valley degree of freedom and spin-orbit interaction (SOI) generally cannot be neglected.
In the absence of a magnetic field, the SOI lifts the fourfold spin-valley degeneracy and splits the four states into two Kramers doublets, $({\rm K},\uparrow)\text{-}({\rm K}',\downarrow)$ and $({\rm K},\downarrow)\text{-}({\rm K}',\uparrow)$, where ${\rm K}$ and ${\rm K}'$ denote the valley indices.
An external magnetic field further induces both spin and orbital Zeeman shifts, the latter being valley dependent.
As a result, the energy spectrum of a realistic CNT quantum dot is more complex than that of the minimal model used in the main analysis~\cite{Kuemmeth2008,Izumida2009,Klinovaja2011}, as schematically shown in Fig.~\ref{fig:magnetic_region}.
The level repulsion between $({\rm K}',\downarrow)$ and $({\rm K},\downarrow)$ 
(or between $({\rm K},\uparrow)$ and $({\rm K}',\uparrow)$)
at finite magnetic fields originates from valley mixing induced by disorder or impurities.

The simplified model is nevertheless applicable in the magnetic-field regime indicated by the red circles in Fig.~\ref{fig:magnetic_region}. 
In this regime, a pair of spin states within the same valley, such as $({\rm K},\uparrow)$ and $({\rm K},\downarrow)$, or equivalently $({\rm K}',\uparrow)$ and $({\rm K}',\downarrow)$, is assumed to dominate the transport, while the other valley-related levels are sufficiently separated in energy. 
The energy splitting between the two relevant states can then be tuned by the external magnetic field and brought into resonance with the torsional-mode energy. 
Thus, the parameter $h$ in the effective model should be understood as the magnetic-field-controlled energy splitting of this isolated spin doublet. 

\subsection{CNT phonon modes}

The relaxation of the torsional mode also deserves a brief comment. Although the twist angle of the fundamental mode is proportional to $\sin(\pi z'/L)$ and therefore vanishes at the two ends of the suspended CNT, this does not imply the absence of mechanical dissipation through the supports. In miniaturized resonators, the quality factor is generally determined by a combination of boundary damping and intrinsic material damping~\cite{Joshi2014}. In particular, support loss originates from time-dependent forces and moments transmitted to the supporting structure, rather than from the endpoint displacement itself~\cite{Joshi2014}. For the present torsional mode, the twist angle vanishes at the ends whereas the torsional strain $\partial_{z'}\theta \propto \cos(\pi z'/L)$ remains finite there, so the suspended CNT can still exchange mechanical energy with the electrodes and surrounding structure. Experimentally, suspended CNT resonators are also known to exhibit dissipation that depends on the clamping configuration, including devices in which the nanotube is held at the electrodes by van der Waals adhesion~\cite{Aykol2014,Kumar2019clamping}. In this work, we therefore treat the torsional-mode relaxation phenomenologically, as an effective description of such environmental dissipation channels.

The interplay among temperature, mode frequency, and mechanical dissipation is also important for the experimental observability of the current. 
The effect is not restricted in principle to temperatures below $0.1\,{\rm K}$: by shortening the CNT, one can increase the fundamental-mode frequency $\omega_0$, thereby raising the torsional-mode energy and enhancing the SRC-induced spin-flip rate. 
At higher temperatures, however, maintaining a large mechanical quality factor $Q$ generally becomes more challenging. 
The resulting increase in mechanical dissipation may therefore suppress the current to a level below experimental detectability.

It is also useful to comment on the continuum description adopted in the present work. Our analysis focuses on the fundamental torsional mode of a suspended CNT, whose wavelength is given by $\lambda = 2L$. For the parameter range considered here, this corresponds to a long-wavelength acoustic phonon, for which a continuum elastic description is generally expected to be reliable~\cite{Suzuura2002,DeMartino2009}.
The validity of continuum elasticity for low-frequency, long-wavelength mechanical modes of CNTs has also been supported by a number of experiments and by comparisons between experiment and continuum modeling~\cite{Xu:2022aa}. 
In this sense, we expect that the main physical picture developed in the present paper, namely resonant spin-to-phonon pumping of a low-energy torsional mode, remains valid beyond the specific parametrization of the continuum model.

In the long-wavelength limit, the low-energy acoustic sector consists of longitudinal, torsional, and flexural modes, whereas the radial breathing mode has a much higher frequency~\cite{Gupta2008}. 
For the parameter set used in Sec.~\ref{sec:method}, the fundamental torsional-mode energy is $\hbar\omega_0 \simeq 280\,\mu{\rm eV}$ for $L=100\,{\rm nm}$ and scales approximately as $1/L$.
Although each mode may produce a resonance, the relatively large quality factors considered here are expected to allow the corresponding peaks to be resolved separately. Since the torsional mode produces the strongest local rotational motion of the lattice, it is expected to couple most strongly to the spin-rotation interaction and therefore to yield the most pronounced current-resonance peak.

A more microscopic treatment based on a calculated CNT phonon spectrum and electron-phonon coupling, for example from atomistic or \textit{ab initio} approaches, would nevertheless be valuable for quantitative refinement. Such an approach could modify the precise mode frequency, the effective coupling strength, and the detailed correspondence between the torsional mode and the underlying CNT phonon branches. However, as long as one remains in the low-energy, long-wavelength regime, these effects are expected to mainly renormalize the quantitative parameters rather than alter the qualitative resonance mechanism itself.

\section{Summary}
\label{sec:summary}

In summary, we have theoretically studied current-induced excitation of a clamped torsional mode in a carbon nanotube (CNT) quantum dot mediated by spin-rotation coupling. The present work extends SRC-mediated Einstein--de Haas actuation from the continuous rigid-body rotation discussed in Ref.~\cite{Izumida2022} to a suspended CNT torsional resonator with a discrete mechanical eigenfrequency.

Our analysis, based on a master-equation approach, shows that when the Zeeman splitting matches the energy of the CNT torsional mode ($h = \hbar \omega_0$), its amplitude is strongly enhanced through resonant pumping. 
Importantly, this mechanism lifts the spin valve effect intrinsic to the antiparallel electrode configuration, giving rise to a measurable leakage current in a regime where transport would otherwise be suppressed.

Numerical estimates for realistic CNT parameters predict a detectable torsional displacement and a current signal on the order of several tens of pA. These signals are well within the sensitivity limits of state-of-the-art nanoelectromechanical measurement techniques. Furthermore, we have confirmed that the mechanism remains robust even in the presence of complex energy spectra due to realistic SOI in CNTs, provided the magnetic field is used as a control parameter.

Our findings indicate that clamped CNT torsional modes provide a promising platform for studying resonant spin-to-mechanical angular-momentum conversion in nanoelectromechanical systems.
The extension of the present mechanism to more transparent junctions, where coherent interference or Kondo correlations may become important, remains an interesting subject for future study.

\begin{acknowledgments}
This work was supported by the National Natural Science Foundation of China (NSFC) under Grant No. 12374126, by the Priority Program of Chinese Academy of Sciences under Grant No. XDB28000000, and by JSPS KAKENHI for Grants (JP24K06951, JP23H01839, JP24H00322, JP21K03414) from MEXT, Japan.
\end{acknowledgments}

\appendix

\section{Hamiltonian formalism of torsional oscillations}
\label{app:Hamiltonian}

In this Appendix, we present a detailed derivation of the quantized torsional modes of a CNT. Within the continuum approximation, the torsional displacement $\theta(z)$ is conjugate to the angular-momentum density $\pi(z)$, and the classical Hamiltonian describing the system is given by
\begin{align}
\label{h_classical_app}
H = \int_0^L dz \, \left[ \frac{1}{2\rho R^2}\pi^2(z) + \frac{GSR^2}{2} ( \partial_{z} \theta(z))^2 \right],
\end{align}
where $\rho$ is the mass density per unit length, $R$ is the radius, $G$ is the shear modulus, and $S=2\pi Rd$ is the cross-sectional area.
For simplicity, we have replaced $z'$ with $z$.
The canonical equations of motion are
\begin{align}
\dot{\theta}(z,t) &= \frac{\delta H}{\delta \pi(z,t)} = \frac{\pi(z,t)}{\rho R^2}, \\
\dot{\pi}(z,t) &= -\frac{\delta H}{\delta \theta(z,t)} = GSR^2 \partial_z^2 \theta(z,t),
\end{align}
leading to the wave equation, Eq.~\eqref{waveequation}.

To quantize the system, we impose the canonical commutation relation $[\hat{\theta}(z), \hat{\pi}(z')] = i\hbar\delta(z-z')$. Under the boundary conditions $\theta(0,t) = \theta(L,t) = 0$, we expand the field operators in terms of normal modes as
\begin{align}
\hat{\theta}(z) &= \sum_{n=1}^\infty \sqrt{\frac{2}{L}} \sin(k_n z) \hat{\theta}_n, \\
\hat{\pi}(z) &= \sum_{n=1}^\infty \sqrt{\frac{2}{L}} \sin(k_n z) \hat{\pi}_n,
\end{align}
where $k_n = n\pi/L$. The mode operators satisfy the discrete commutation relation $[\hat{\theta}_n, \hat{\pi}_{n'}] = i\hbar\delta_{nn'}$. Substituting these expansions into Eq.~(\ref{h_classical_app}) and utilizing the orthogonality of the sine functions, the Hamiltonian is diagonalized as
\begin{align}
H = \sum_{n=1}^\infty \left[ \frac{1}{2\rho R^2} \hat{\pi}_n^2 + \frac{\rho R^2 \omega_n^2}{2} \hat{\theta}_n^2 \right],    
\end{align}
where $\omega_n = c k_n$ is the eigenfrequency of the $n$-th mode.

Next, we introduce the bosonic creation and annihilation operators $\hat{a}_n^\dagger$ and $\hat{a}_n$:
\begin{align}
\hat{a}_n &= \sqrt{\frac{\rho R^2 \omega_n}{2\hbar}} \left( \hat{\theta}_n + i \frac{\hat{\pi}_n}{\rho R^2 \omega_n} \right), \\
\hat{a}_n^\dagger &= \sqrt{\frac{\rho R^2 \omega_n}{2\hbar}} \left( \hat{\theta}_n - i \frac{\hat{\pi}_n}{\rho R^2 \omega_n} \right).
\end{align}
These operators satisfy $[\hat{a}_n, \hat{a}_{n'}^\dagger] = \delta_{nn'}$, and the Hamiltonian takes the standard second-quantized form $H = \sum_{n} \hbar \omega_n (\hat{a}_n^\dagger \hat{a}_n + 1/2)$.
In the Heisenberg picture, the time-dependent torsional operator for the fundamental mode ($n=1$) is expressed as
\begin{align}
\hat{\theta}_1(t) &= \sqrt{\frac{\hbar}{2\rho R^2 \omega_1}} (\hat{a}_1 e^{-i\omega_1 t} + \hat{a}_1^\dagger e^{i\omega_1 t}), \\
\hat{\pi}_1(t) &= -i \sqrt{\frac{\hbar \rho R^2 \omega_1}{2}} (\hat{a}_1 e^{-i\omega_1 t} - \hat{a}_1^\dagger e^{i\omega_1 t}).
\end{align}
The former equation corresponds to the time evolution of the operator $\hat{\theta}$ given in Eq.~(\ref{eq:theta2}) of the main text.

\bibliography{ref}

\end{document}